\documentclass{Interspeech2024}
\usepackage{multirow}
\usepackage{cite}
\usepackage{setspace}
\usepackage{caption}
\usepackage{subcaption}
\usepackage{algpseudocode}
\usepackage{algorithm}
\usepackage{colortbl}
\usepackage{tabularx}
\usepackage{threeparttable,booktabs}
\usepackage[dvipsnames]{xcolor}

\algtext*{EndFor}
\makeatletter
\def\bstctlcite{\@ifnextchar[{\@bstctlcite}{\@bstctlcite[@auxout]}}
\def\@bstctlcite[#1]#2{\@bsphack
  \@for\@citeb:=#2\do{%
    \edef\@citeb{\expandafter\@firstofone\@citeb}%
    \if@filesw\immediate\write\csname #1\endcsname{\string\citation{\@citeb}}\fi}%
  \@esphack}
\makeatother 

\algnewcommand\algorithmicto{%
    \textbf{to}}
\algdef{SE}[FORINPARALLEL]{ForParallel}{EndForParallel}{\algorithmicfor\ \algorithmicto}{\algorithmicend\ \algorithmicfor}
\algtext*{EndForParallel}
\interspeechcameraready

\title{Towards Effective and Efficient Non-autoregressive Decoding \\Using Block-based Attention Mask }

\name[affiliation={1,2*}]{Tianzi}{Wang}
\name[affiliation={2\dagger}]{Xurong}{Xie}
\name[affiliation={1}]{Zhaoqing}{Li}
\name[affiliation={3}]{Shoukang}{Hu}
\name[affiliation={1}]{Zengrui}{Jin}
\name[affiliation={1}]{Jiajun}{Deng}
\name[affiliation={1}]{Mingyu}{Cui}
\name[affiliation={1}]{Shujie}{Hu}
\name[affiliation={4}]{Mengzhe}{Geng}
\name[affiliation={1}]{Guinan}{Li}
\name[affiliation={1}]{Helen}{Meng}
\name[affiliation={1}]{Xunying}{Liu}

\address{
  $^1$The Chinese University of Hong Kong, \thanks{$*$ Work done while at Institute of Software, Chinese Academy of Sciences during the internship}
  $^2$Institute of Software, Chinese Academy of Sciences\thanks{$\dagger$ \text{ Corresponding author}}\\
  $^3$Nanyang Technological University,  
  $^4$National Research Council Cananda}
\email{twang@se.cuhk.edu.hk, xurong@iscas.ac.cn, xyliu@se.cuhk.edu.hk}

\keywords{Speech recognition, Non-autoregressive decoder, Autoregressive decoder, 
Beam search, Conformer}

\begin{document}
\bstctlcite{IEEEexample:BSTcontrol}
\maketitle

\begin{abstract}
This paper proposes a novel non-autoregressive (NAR) block-based Attention Mask Decoder (AMD) that flexibly balances performance-efficiency trade-offs for Conformer ASR systems. AMD performs parallel NAR inference within contiguous blocks of output labels that are concealed using attention masks, while conducting left-to-right AR prediction and history context amalgamation between blocks. A beam search algorithm is designed to leverage a dynamic fusion of CTC, AR Decoder, and AMD probabilities. Experiments on the LibriSpeech-100hr corpus suggest the tripartite Decoder incorporating the AMD module produces a maximum decoding speed-up ratio of 1.73x over the baseline CTC+AR decoding, while incurring no statistically significant word error rate (WER) increase on the test sets. When operating with the same decoding real time factors,  statistically significant WER reductions of up to 0.7\% and 0.3\% absolute (5.3\% and 6.1\% relative) were obtained over the CTC+AR baseline.
\end{abstract}

\section{Introduction}
State-of-the-art Transformer based automatic speech recognition (ASR) systems represented by, for example, Conformer Encoder-Decoder models with joint CTC and attention costs \cite{watanabe2017hybrid,guo2021recent,kim2017joint,sainath2019two,yao2021wenet}, are often based on an autoregressive (AR) Transformer Decoder architecture. The resulting AR model inference follows a monotonic and left-to-right decoding strategy, which predicts one output token at a time. However, this limits the potential for model inference parallelization during beam search decoding for practical application scenarios that are not only performance-critical but also efficiency-sensitive. 

Several methods have been studied to optimize the inference speed of ASR models with Encoder-Decoder architecture \cite{gandhi2023distil, lilossless,qian2019binary,wang2015small,he2019streaming, ng2021pushing, wang2021streaming, gao2022paraformer,higuchi2020mask,higuchi2021improved,chan2020imputer,chen2020non,chi2020align, chen2021align,sudo20224d,tian2022hybrid,wang2022deliberation}. One general solution is to adopt non-autoregressive (NAR) Transformers based Decoder designs. NAR Transformers provide more powerful parallelization than AR architectures to improve ASR decoding efficiency, and have gained considerable attention in a series of prior researches in recent years \cite{higuchi2020mask, ng2021pushing, chi2020align, wang2021streaming, gao2022paraformer}. These include, but not limited to, the following categories: a) mask-based NAR \cite{higuchi2020mask,higuchi2021improved,chan2020imputer,chen2020non} learn to fill the randomly masked training label tokens conditioned on the unmasked ones; b) alignment-refinement based approaches \cite{chi2020align, chen2021align} that aim to refine the initial CTC produced alignment by injecting noisy labels extracted from, for example, auto-encoders in the Align-Refine method \cite{chi2020align}, or noise perturbed Encoder alignment posteriors in the Align-Denoise approach \cite{chen2021align}; c) integrate-and-fire based approaches \cite{gao2022paraformer} which learn to align speech and labels using continuous integrate-and-fire modules; and d) hybrid AR+NAR Decoders that exploit their complementarity in combination \cite{sudo20224d,tian2022hybrid,wang2022deliberation}. 

Efforts on developing high-performance and low-latency NAR based ASR models require several key challenges to be addressed. First, NAR models’ intrinsic lack of monotonic sequence modelling constraints leads to their large performance gap against state-of-the-art ASR systems based on AR designs. 
Despite recent attempts to mitigate such modelling deficiency of NAR ASR systems using, for example, Mask-CTC and its improved variants \cite{higuchi2020mask, higuchi2021improved}, or iterative alignment-refinement approaches \cite{chi2020align, chen2021align}, their performance gap against AR counterparts still exists. Second, there is a notable lack of effective and efficient one-pass beam search decoding algorithm that are purpose-designed for NAR Decoders and their combination with CTC and AR ones. Prior researches in this direction either deployed standalone NAR Decoders in a later rescoring pass within a multi-pass decoding framework \cite{sudo20224d}, while the first pass recurrent neural network transducer (RNN-T) decoding serves to produce an initial set of hypotheses, or vice versa when the NAR Decoders are used in the initial N-best generation before AR rescoring \cite{wang2022deliberation}. 

To this end, this paper proposes a novel non-autoregressive block-based attention mask decoder (AMD) that flexibly balances performance-efficiency trade-offs for Conformer ASR systems. The AMD decoder leverages both parallel NAR inference and monotonic left-to-right AR prediction. Parallel NAR inference is performed within contiguous blocks of output labels that are concealed using attention masks, while monotonic and left-to-right AR prediction and history context amalgamation are conducted between blocks. The AMD based NAR decoder is jointly trained with the standard CTC module and attention-based AR Transformer. A beam search algorithm is designed to leverage a dynamic fusion of CTC, AR Decoder and AMD probabilities. In addition to fixed size attention-masking blocks during NAR inference, mixed size blocks were also explored to facilitate cold start monotonic inference (block size $=$ 1) for the initial $N$ labels of each speech segment, before switching to parallel label prediction (block size $>$ 1) for the remaining labels. 

Experiments were conducted on the benchmark LibriSpeech-100 dataset using smaller Conformer ASR models trained on the 100-hr data only, or larger ones \cite{chang2021exploration} further incorporating WavLM SSL features and 960-hr data based pre-training. The tripartite Decoder incorporating the additional AMD module produces a maximum decoding speed up ratio of 1.73x over the baseline CTC+AR decoding, while incurring no statistically significant word error rate (WER) increase on the test sets on average. When operating with the same decoding real time factors,  statistically significant WER reductions of up to 0.7\% and 0.3\% absolute (5.2\% and 6.1\% relative) were obtained over the CTC+AR baseline.

The main contributions of this paper are as follows: 
\begin{itemize}
\item This paper presents a novel block-based attention mask decoder (AMD) that flexibly balances performance-efficiency trade-offs for Conformer ASR systems in recognition time. For the first time, this NAR decoder allows: 1) decoding time speed up via non-autoregressive, parallel inference without increasing ASR WERs; and 2) statistically significant WER reductions over AR decoders when operating with the same decoding real time factors.  In contrast, large performance degradation were often observed in prior researches when alternative forms of NAR Decoders \cite{higuchi2020mask,higuchi2021improved,chen2021align}, e.g. Mask-CTC, were used, or failed to provide the WERs at equivalent real-time factors for fair comparison \cite{gao2022paraformer}. 

\item A novel beam search algorithm is designed to leverage a dynamic fusion of CTC, AR Decoder and AMD probabilities. In contrast, prior researches have largely deployed NAR Decoders in a more time-consuming multi-pass decoding framework \cite{sudo20224d, wang2022deliberation, song2021non, pmlr-v139-qi21a, liang-etal-2022-janus}. 




\end{itemize}

\begin{figure}[t]
    \centering
    \includegraphics[width=0.8\linewidth]{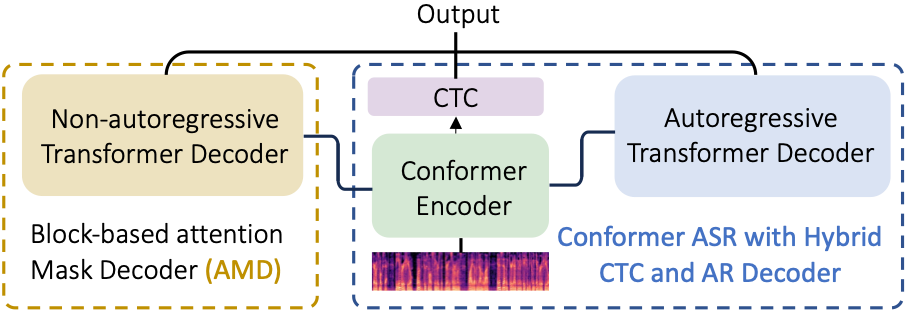}
    \vspace{-3mm}
    \caption{The proposed Conformer ASR system architecture using a tripartite Decoder that includes the proposed non-autoregressive block-based attention mask decoder (AMD) (left, dashed yellow line) in addition to the CTC module (centre, purple) and attention-based AR Decoder (right, dashed blue line). }
    \vspace{-7mm}
    \label{fig:main}
\end{figure}

\vspace{-3mm}
\section{Hybrid Attention Encoder-Decoder Based ASR System}
A hybrid CTC-attention Encoder-Decoder (AED) based ASR system considered in this paper consists of three parts \cite{watanabe2017hybrid}: a shared Conformer-based Encoder, a CTC Decoder and an attention-based AR Decoder. An interpolated CTC and AR cost function is used for model training: 
\vspace{-1mm}
\begin{equation}
\vspace{-1mm}
\mathcal{L}_{\text{c-a}}=\gamma_{1}\mathcal{L}_{\text{CTC}} + \gamma_{2}\mathcal{L}_{\text{AR}}
\label{eq:hybrid2}
\end{equation}
where $\gamma_1$ and $\gamma_2$ are the weights for CTC and AR Decoder during training respectively. The decoding stage follows an label synchronous autoregressive beam search \cite{watanabe2017hybrid} to find the hypothesis by taking into account both the CTC and AR Decoder probabilities.
At the $i$-th step of the decoding process, the score of the partially decoded hypothesis $\boldsymbol{h}_{\leq i}$ is given by:
\vspace{-1mm}
\begin{equation}
\vspace{-1mm}
\alpha_{\text{c-a}}(\boldsymbol{h}_{\leq i})=\lambda_{1}\alpha_{\text{CTC}}(\boldsymbol{h}_{\leq i}) + \lambda_{2}\alpha_{\text{AR}}(\boldsymbol{h}_{\leq i})
\label{eq:c-a}
\end{equation}
where $\lambda_1$ and $\lambda_2$ are the weights for CTC and AR Decoder during decoding respectively.
The CTC score $\alpha_{\text{CTC}}$ is expressed as the cumulative probability of all token sequences sharing $\boldsymbol{h}_{\leq i}$ as the common history context given by 
\vspace{-1mm}
\begin{equation}
\vspace{-1mm}
\alpha_{\text{CTC}}(\boldsymbol{h}_{\leq i})=\text{log}P_{\text{CTC}}(\boldsymbol{h}_{\leq i}, \cdots|\boldsymbol{\mathcal{X}})
\label{eq:ctc_p}
\end{equation}
where $\boldsymbol{\mathcal{X}}$ denotes the Encoder outputs. Meanwhile, the AR Decoder probability of the output token $y_i$ is conditioned on $\boldsymbol{h}_{<i}$ and $\boldsymbol{\mathcal{X}}$ as $P_{AR}(y_i|\boldsymbol{h}_{<i},\boldsymbol{\mathcal{X}})$.
The AR Decoder score $\alpha_{\text{AR}}(\boldsymbol{h}_{\leq i})$ is given by:
\vspace{-1mm}
\begin{equation}
\alpha_{\text{AR}}(\boldsymbol{h}_{\leq i})=\text{log}\prod_{j=1}^iP_{\text{AR}}(y_{j}|\boldsymbol{h}_{<j},\boldsymbol{\mathcal{X}})
\label{eq:ar_p}
\vspace{-2mm}
\end{equation}
\section{Block-Based Attention Mask Decoder}
The proposed tripartite Decoder integrates a novel non-autoregressive block-based attention mask decoder (AMD) (Fig. \ref{fig:main}, left, dashed yellow line) alongside the standard CTC module (Fig. \ref{fig:main}, center, purple) and attention-based AR Decoder (Fig. \ref{fig:main}, right, dashed blue line). 
\begin{figure}[t]
    \centering
    \includegraphics[width=\linewidth]{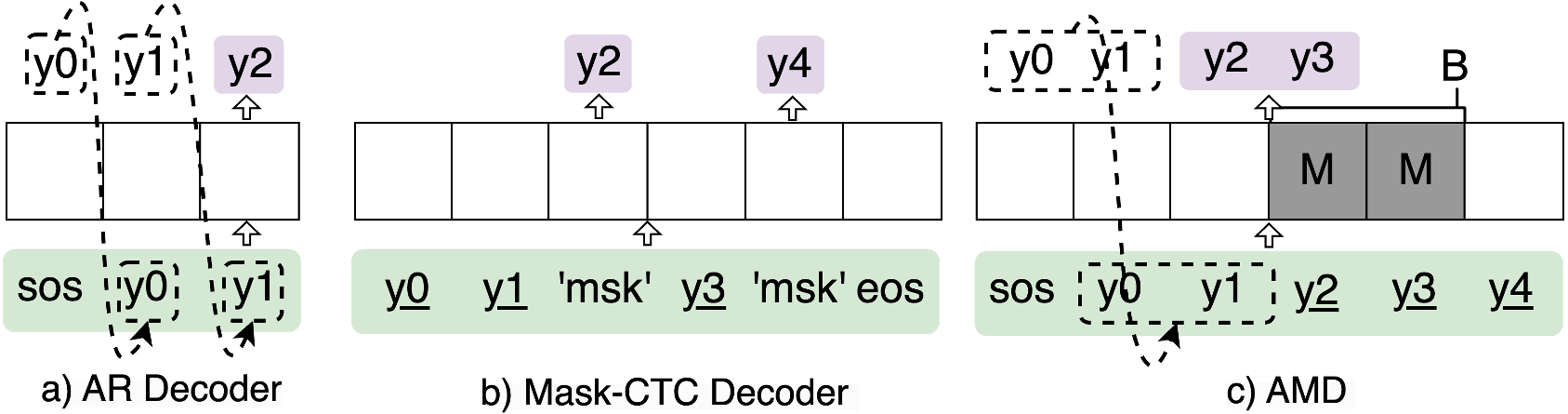}
    \vspace{-6mm}
    \caption{Inference using: a) an AR Decoder, b) a Mask Prediction Decoder;  c) the proposed AMD. `msk' refers to input mask tokens, \text{\rm\colorbox{gray}{M}} is the attention mask within contiguous blocks for parallel inference via AMD, `B' refers to the block size. Tokens highlighted in green background denote Decoder inputs at the current inference step. Tokens highlighted in purple denote predicted tokens at current step. Tokens in dashed boxes in a) and c) represent those from the previous inference step. Tokens with ``\_" denote those obtained from CTC prediction. }
    \label{fig:attn}
    \vspace{-7mm}
\end{figure}
\subsection{Block-Based Attention-Masking Decoder}
\label{sec:3}

Fig. \ref{fig:attn} illustrates example inference using either: a) a AR Decoder; b) a Mask Prediction Decoder;  or c) the proposed AMD.
The AR decoder in Fig. \ref{fig:attn} assigns the probability of each output token in a strictly autoregressive left-to-right and monotonic manner. The NAR Mask-CTC Decoder \cite{higuchi2020mask} performs parallel inference over multiple input tokens that are concealed using special “msk” symbols, e.g. ${y_2}$ and $y_4$, while the sequential constraint between the masked input tokens ignored. It is also possible to restrict Mask-CTC prediction to be applied only to a smaller subset of the output tokens that are carefully selected using CTC scores during recognition time, in order to mitigate the performance loss when used to predict all the tokens \cite{higuchi2020mask} (See contrast between Sys. 2 and 3, Tab. \ref{tab:greedy_search}). 

In contrast, the proposed AMD model leverages both parallel NAR inference and AR prediction as illustrated in Fig. \ref{fig:attn}(c). Parallel NAR inference is performed within each contiguous block of output tokens concealed via block-based attention masks \colorbox{gray}{M}, e.g. ${y_2}$ and ${y_3}$, while sequential AR prediction and history context accumulation are made left-to-right between the blocks. To prevent information leakage in AMD, the same input token position being attention-masked also requires its respective token embedding layer outputs to be set as zero. In this paper, AMD is used to predict all tokens without any subset filtering as in Mask-CTC prediction. 

Given an input token sequence  $\boldsymbol{h}$ and and the Encoder outputs $\boldsymbol{\mathcal{X}}$, the AMD probability $P_{\text{AMD}}(y_j|\boldsymbol{h}, \boldsymbol{\mathcal{X}})$ of the $j$-th token which are in an attention-mask block, $j$-th $\in [i, i+B-1]$,
\vspace{-1mm}
\begin{equation}
P_{\text{AMD}}(y_{j}|\boldsymbol{h}, \boldsymbol{\mathcal{X}})=P_{\text{AMD}}(y_j|\boldsymbol{h}_{<i},\boldsymbol{h}_{>i+B-1}, \boldsymbol{\mathcal{X}})
\label{eq:amd_p}
\vspace{-1mm}
\end{equation}
where $B$ is the size of the attention-mask block. 

Ground truth text labels are used as the input during AMD training. In order to improve the AMD model’s generalization to all the tokens in the training data, and also different inference block sizes (including using mixed sized blocks of Sec. \ref{sec:3.2}), diversity is injected into the training process by repeatedly performing inference over all the tokens of each sentence using attention-mask blocks of 4 different sizes that are randomly sampled within the range of [1, $L$], where $L$ is the length of each sentence. The AMD training loss is defined as the probabilities accumulated over all tokens in a sentence using a block size $B_n$, before being marginalized over different block sizes:
\vspace{-2mm}
\begin{equation}
\label{eq:3}
\mathcal{L}_{\text{AMD}}=-\sum_{n=1}^4\text{log}\prod_{j=1}^L P_{\text{AMD}}(y_{j}|\boldsymbol{h}_{<i},\boldsymbol{h}_{>i+B_n-1}, \boldsymbol{\mathcal{X}})
\vspace{-2mm}
\end{equation}
The 3-way interpolated $\mathcal{L}_{\text{CTC}}$, $\mathcal{L}_{\text{AR}}$, and $\mathcal{L}_{\text{AMD}}$ losses was used to form a tripartite loss:
\vspace{-2mm}
\begin{equation}
\label{eq:loss-cma}
\mathcal{L}_{\text{c-m-a}}=\gamma_{1}\mathcal{L}_{\text{{CTC}}}+\gamma_{2}\mathcal{L}_{\text{{AR}}}+\gamma_{3}\mathcal{L}_{\text{{AMD}}}
\vspace{-2mm}
\end{equation}
where $\gamma_1$,  $\gamma_2$ and $\gamma_3$ are the weights for CTC, AR Decoder and AMD during training respectively. This combined loss is used to train the Conformer ASR model featuring the additional AMD module in Fig. \ref{fig:main}.

\subsection{Beam Search Using AMD}
\label{sec:3.2}
\begin{figure}[t]
    \centering
    \includegraphics[width=\linewidth]{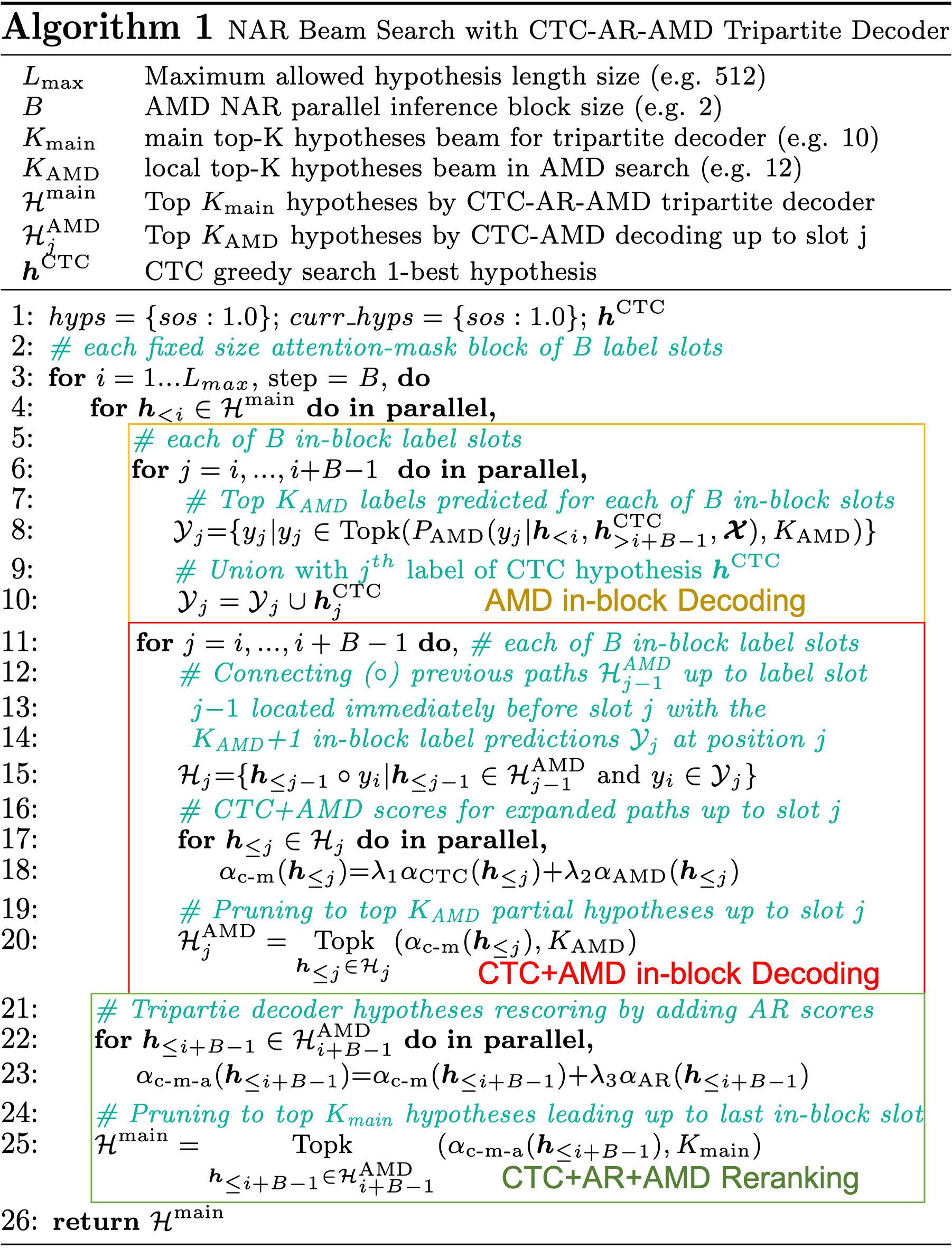}
    \label{fig:algo}
    \vspace{-7mm}
\end{figure}

AMD Decoding differs from the above training procedure mainly by replacing the ground truth input labels to be block attention-masked in training using the label hypotheses obtained from CTC greedy search. The AMD probabilities are dynamically fused with the CTC and AR Decoder scores in a novel beam search algorithm (Algorithm 1) tailored designed for AMD in this paper. 

In addition to the above mentioned NAR style parallel inference within each block (line 5-8, Algorithm 1), there are several further important details of this algorithm to note: 

\noindent\textbf{(1)} Left-to-right sequential AR prediction requires commensurate history context accumulation between the moving blocks. This is achieved by connecting the existing partial sequence hypotheses leading up to a $(j{-}1)$-th label slot, immediately before the current in-block $j$-th label being predicted, with the top-K output token labels for the $j$-th label (line 11-15). 

\noindent\textbf{(2)} To improve hypotheses coverage, the CTC greedy search 1-best hypothesis is progressively merged with the AMD decoding outputs (line 9-10).

\noindent\textbf{(3)} To ensure efficiency during AMD search, top-K pruning is initially applied locally to the output tokens for each in-block slot using AMD scores alone (line 8, via $K_\text{AMD}$). It is further applied to the iteratively expanded partial sequence hypotheses using interpolated CTC+AMD scores with $\lambda_1$ and $\lambda_2$ as the weights of CTC and AMD respectively (line 16-18, via $K_\text{AMD}$)\footnote{CTC and AMD scores weights empirically set as 0.3 and 0.3}, and finally to the tripartite CTC+AR+AMD decoder scores re-ranked hypotheses with $\lambda_3$ as the weight for AR Decoder(line 21-23, via $K_\text{main}$)\footnote{CTC-AMD and AR scores weighs empirically set as 0.6 and 0.4}. 
The CTC score can be calculated efficiently following \cite{watanabe2017hybrid}, and the AMD score for each partial hypothesis sequence $\boldsymbol{h}_{\leq i}$, $\alpha_{\text{AMD}}(\boldsymbol{h}_{\leq i})$ is 
\vspace{-3mm}
\begin{equation}
\alpha_{\text{AMD}}(\boldsymbol{h}_{\leq i})=\text{log}\prod_{j=1}^iP_{\text{AMD}}(y_j|\cdot)
\label{eq:amd_a}
\vspace{-3mm}
\end{equation}
where the AMD probability $P_{\text{AMD}}(y_j|\cdot)$ is defined in Eqn. (\ref{eq:amd_p}).

To achieve more flexible balance between performance and efficiency, in addition to using fixed size attention-masking blocks during NAR inference, \textbf{mixed size blocks} were also explored to allow cold start monotonic inference (block size $B{=}1$) for the initial N labels of each speech segment to be more slowly built-up, before switching to faster parallel label prediction ($B{>}1$) for the remaining labels in the same sentence. 

\section{Experiments}
\noindent
\noindent
\textbf{Experiment Setup and Baseline Systems: }
Experiments were conducted using {\bf (1)} smaller Conformer end-to-end models \footnote{feedforward dim = 1024, \# attention head = 4,  attention head dim = 256,  convolution kernel size = 15.} with 80-dim log-mel filter-banks and 3-dim pitch as input, and {\bf (2)} larger Conformer models \footnote{feedforward dim = 2048, \# attention head = 8,  attention head dim = 512,  convolution kernel size = 15.} further incorporating WavLM  \cite{chen2022wavlm} SSL features and 960-hr data based pre-training.
All models are of 2 Convolution blocks followed by 12 Conformer blocks for encoder, and 6 Transformer decoder blocks. The decoder output vocabulary comprised 5,000 byte pair encoding (BPE) tokens derived from the transcripts of respective training sets.
For all models with an CTC+AR Decoder, an interpolated CTC+AED (3:7) loss function was used for training. Speed perturbation \cite{ko2015audio} and SpecAugment \cite{Park2019} were also applied.
Real time factors (RTFs) of the smaller Conformer models were measured on an Intel Xeon 5317, while RTFs of the larger ones were measured on an NVIDIA A40.
Matched pairs sentence-segment word error (MAPSSWE) \cite{gillick1989some} based statistical significance test was performed at a significance level $\alpha=0.05$. 

\begin{table}[t]
\scriptsize
\setlength\tabcolsep{1.5pt}
\caption{Performance of {\bf (a)} greedy search baseline with CTC + autoregressive (AR) Decoder (Sys. 1, 13); {\bf (b)} Mask-CTC Decoder \cite{higuchi2020mask} implemented in the form of CTC incorporated with masked language model (MLM), with approx. 10\% (Sys. 2, selected acc. CTC scores) and all (Sys. 3) of the tokens refilled by MLM Decoder; {\bf (c)} bipartite CTC + AMD based Decoder (Sys. 5-6) ; {\bf (d)} the proposed tripartite Decoder with fixed decoding block size (Sys. 4-9, 14-16), and with variable size decoding (Sys. 10-12). 
``N-B'' denotes first $N$ tokens are decoded in an AR manner, followed by non-AR decoding of remaining tokens with $\text{block size}=B$. ``1-best'' stands for the 1-best WER. \colorbox{yellow!20}{$\dagger$} and \colorbox{yellow!20}{$\ddagger$} denote no statistically significant WER difference over Sys. 1 and 13 respectively on the average (Ave.) WER, \colorbox{blue!20}{$\diamond$} and \colorbox{blue!20}{$\triangleright$} denote RTF similar to Sys. 1 and 13 respectively. 
}
\vspace{-3mm}
\label{tab:greedy_search}
\resizebox{\linewidth}{!}{
\begin{tabular}{c|c|c|c|c|ccc|c}
\hline\hline
\multirow{2}{*}{Sys.} & \multirow{2}{*}{Encoder}  & \multirow{2}{*}{Decoder}   & \multirow{2}{*}{Weights}        & \multirow{2}{*}{\begin{tabular}[c]{@{}c@{}}Block\\Size\end{tabular}}& \multicolumn{3}{c|}{1-best}   & \multirow{2}{*}{RTF} \\ \cline{6-8}
 &    &     &  & & clean &  other & Ave. &            \\ \hline
1 &   \multirow{11}{*}{\begin{tabular}[c]{@{}c@{}}Conformer\\(100-hr)\end{tabular}}    & CTC+AR & 0.7:0.3 & 1 & 7.8 & 18.9 & 13.3 & 0.26 \\
\cline{1-1}\cline{3-5}
2 &   & CTC+MLM \cite{higuchi2020mask} ($10\%$)  & - & - & 7.5 &  20.6  &  14.0  &  0.07 \\ \cline{1-1} \cline{3-5}
3   &  & CTC+MLM \cite{higuchi2020mask} ({\it all}) & -   & -            &   10.7    &  21.1   &  15.9 &  0.10      \\\cline{1-1}\cline{3-5}
4    &                    & \multirow{2}{*}{CTC+AMD}  & \multirow{2}{*}{0.5:0.5}        & 1                           & 7.6 
& 18.7
& 13.2     & 0.26                 \\ \cline{1-1} 
5   &                                     &       &                                 & 2                           & 8.9        & 20.9       & 14.4     & 0.17                 \\ \cline{1-1}\cline{3-5}
6    &                    &  \multirow{7}{*}{CTC+AR+AMD} &\multirow{7}{*}{\begin{tabular}[c]{@{}c@{}}0.4:\\0.3:0.3\end{tabular}} & 1                           & 6.7        & 17.9       & 12.3    & 0.41                 \\ \cline{1-1} 
7    &                    &                          &                                 & 2                           & 7.1       & 18.8       & 12.9     & 0.31                 \\ \cline{1-1} 
8     &                                 &      &                                 & 4                           & 7.3        & 19.1       & \cellcolor{yellow!20}{13.2$^\dagger$}     & 0.17        \\ \cline{1-1}
9     &                          &        &                                 & 8                           & 7.5        & 19.4       & \cellcolor{yellow!20}{13.4$^\dagger$}  & 0.15        \\ \cline{1-1} 
10        &            &             &                                 & 10-2                           & 7.2        & 18.6       &   12.9 &      \cellcolor{blue!20}{0.27$^\diamond$}       \\ \cline{1-1}
11     &                &               &                                 & 10-4                           & 7.2        & 18.7       &  12.9 &            \cellcolor{blue!20}{0.26$^\diamond$}   \\ \cline{1-1} 
12      &            &                    &                                 & 30-8                           & 7.0        & 18.3       & 12.6  &      \cellcolor{blue!20}{0.27$^\diamond$}           \\ \hline \hline
13 & \multirow{4}{*}{\begin{tabular}[c]{@{}c@{}}WavLM feat. \\+ Conformer\\(960-hr pt.+\\100-hr ft.)\cite{chang2021exploration} \end{tabular}}  & CTC+AR$*$ & 0.7:0.3 & 1 & 3.8 & 6.1 & 4.9 & 0.069 \\ \cline{1-1}\cline{3-5}
14 & & \multirow{3}{*}{CTC+AR+AMD$*$} & \multirow{3}{*}{\begin{tabular}[c]{@{}c@{}}0.4:\\0.3:0.3\end{tabular}} & 1                           &  3.2      &  5.5     &  4.3 & 0.102               \\ \cline{1-1} 
15      &                     &        &                                 & 4                           & 3.4        &    5.8   &  4.6   &  \cellcolor{blue!20}{0.064$^\triangleright$}              \\ \cline{1-1}
16      &        &      &                                 & 16                           &   3.7      &   6.1  &   \cellcolor{yellow!20}{4.9$^\ddagger$} &    0.048      \\ 
\hline\hline
\end{tabular}}
\begin{tablenotes}
       \item * The CTC and AR Decoder parameters were initialized with the pre-trained model \cite{chang2021exploration} and were frozen during fine-tuning. The AMD Decoder parameters were initialized with \cite{chang2021exploration}  and fine-tuned to LibriSpeech-100hr.
     \end{tablenotes}
\vspace{-6mm}
\end{table}

\begin{table}[t]
\scriptsize
\setlength\tabcolsep{1.5pt}
\caption{Performance of {\bf (a)} baseline CTC+AR beam search (Sys. 1, 9), {\bf (b)} the proposed tripartite Decoder with fixed decoding block (Sys. 2-5, 10-11), and {\bf (c)} with variable size decoding (Sys. 6-8, 12). 
Beam size for all systems was set to 10. ``Oracle'' denotes the Oracle WER obtained from 100-best hypotheses. \colorbox{yellow!20}{$\dagger$} and \colorbox{yellow!20}{$\ddagger$} denote no significant statistically significant WER difference over Sys. 1 and 9 respectively on the average (Ave.) WER, \colorbox{blue!20}{$\diamond$} and \colorbox{blue!20}{$\triangleright$}  denote RTF similar to Sys. 1 and 9 respectively. Other naming conventions follow Tab. \ref{tab:greedy_search}. }
\label{tab:beam_search}
\resizebox{\linewidth}{!}{
\begin{tabular}{c|c|c|c|c|ccc|cc|c}
\hline\hline
\multirow{2}{*}{Sys.} & \multirow{2}{*}{Encoder} & \multirow{2}{*}{Decoder}       & \multirow{2}{*}{Weights}     & \multirow{2}{*}{\begin{tabular}[c]{@{}c@{}}Block\\Size\end{tabular}}   & \multicolumn{3}{c|}{1-best}                                              & \multicolumn{2}{c|}{Oracle}      & \multirow{2}{*}{RTF} \\ \cline{6-10}
    &  &  & &  & \multicolumn{1}{c|}{clean} & \multicolumn{1}{c|}{other} & Ave. & \multicolumn{1}{c|}{clean} &  other &              \\ \hline
1 &  \multirow{7}{*}{\begin{tabular}[c]{@{}c@{}}Conformer\\(100-hr)\end{tabular}}   & CTC+AR & 0.7:0.3 & 1 & 6.8 & 17.5 & 12.1 & 4.3 & 13.6 & 1.19                 \\ \cline{1-1} \cline{3-5} 
2                     & &  \multirow{7}{*}{\begin{tabular}[c]{@{}c@{}}CTC+AR\\+AMD\\\end{tabular}}  & \multirow{7}{*}{\begin{tabular}[c]{@{}c@{}}0.4:\\0.3:0.3\end{tabular}} & 1                           & 6.5        & 17.4       & 11.9    & 4.4        & 13.7     & 1.33                 \\ \cline{1-1} 
3                     &&                              &                              & 2                           & 6.7        & 17.9       & 12.3    & 4.9        & 15.9       & 0.74                 \\ \cline{1-1} 
4                     &&                              &                              & 4                           & 6.9        & 18.3       & 12.6    & 5.2        & 15.6       & 0.45                 \\ \cline{1-1} 
5                     &&                              &                              & 8                           & 7.1       & 18.6       & 12.8    & 5.4        & 15.8       & 0.31                 \\ \cline{1-1} 
6                     &&                              &                              & 5-2                     & 6.7        & 17.8       & \cellcolor{yellow!20}{12.2$^\dagger$}    & 4.8        & 14.7       & 0.75                 \\ \cline{1-1}  
7                     &&                              &                              & 10-2                    & 6.6        & 17.7       & \cellcolor{yellow!20}{12.1$^\dagger$}    & 4.7        & 14.6       & 0.84                 \\ \cline{1-1} 
8                    &&                              &                              & 50-2                    & 6.5        & 17.5       & \cellcolor{yellow!20}{12.0$^\dagger$}    & 4.4        & 13.9       & \cellcolor{blue!20}{1.21$^\diamond$}
\\ \hline\hline
9 &\multirow{4}{*}{\begin{tabular}[c]{@{}c@{}}WavLM feat. \\+ Conformer\\(960-hr pt.+\\100-hr ft.)\cite{chang2021exploration} \end{tabular}}  & CTC+AR & 0.7:0.3 & 1 & 3.1 & 5.4 & 4.2 & 2.1 & 3.8 & 0.124 \\ \cline{1-1} \cline{3-5} 
10 & &  \multirow{3}{*}{\begin{tabular}[c]{@{}c@{}}CTC+AR\\+AMD\end{tabular}}  &  \multirow{3}{*}{\begin{tabular}[c]{@{}c@{}}0.4:\\0.3:0.3\end{tabular}} & 1         &      3.0            &   5.3     &   \cellcolor{yellow!20}{4.1$^\ddagger$}    &   2.1  &          3.7 & 0.178      \\ \cline{1-1} 
11                     &   &        &        & 8        &     3.3    &    5.5   &  \cellcolor{yellow!20}{4.3$^\ddagger$}   & 2.7  &      4.5 & 0.081   \\ \cline{1-1} 
12                     &   &        &        & 10-2        &     3.0    &    5.4   &  \cellcolor{yellow!20}{4.1$^\ddagger$}   &  2.3 &   4.1  & \cellcolor{blue!20}{0.119$^\triangleright$}   \\ \cline{1-1} 
\hline\hline
\end{tabular}}
\vspace{-2mm}
\end{table}

\noindent
\textbf{Experiments Using Greedy Search:}  
Tab. \ref{tab:greedy_search} presents the performance of AMD tripartite and baseline CTC+AR Decoders using greedy search.
Following trends can be observed: 
\textbf{(a)} Bipartite CTC+AMD (Sys. 4-5) and tripartite CTC+AR+AMD (Sys. 6-12) Decoders achieve lower WER but slower speeds compared to Mask-CTC with an MLM decoder (Sys. 2-3), which refills tokens partially.
\textbf{(b)} The CTC+AMD Decoder achieves 1-best WER and RTF similar to the CTC+AR baseline (Sys. 4 \textit{vs.} 1). 
\textbf{(c)} The proposed tripartite Decoder achieves a maximum speed-up of 1.73x over the baseline CTC+AR Decoder, with no significant WER changes on the averaged WERs across ``clean'' and ``other'' test sets using fixed-size decoding (Sys.9 \textit{vs.} 1). 
Mixed-size decoding allows more accurate AR inference to be used for the first $N$ tokens of each sentence, before switching to faster NAR parallel prediction. By doing so performance is weighed more against efficiency.  
Notably, mixed-size decoding systems with RTFs similar to the CTC+AR baseline (Sys. 10-12) achieve statistically significant WER reductions of up to 0.7\% absolute (5.3\% relative) (Sys. 12 \textit{vs.} 1).
\textbf{(d)} Similar trends are observed on larger models incorporating WavLM SSL features and 960-hr pre-training. 
The tripartite Decoder with the AMD module (Sys. 14-16) achieves a maximum speed-up of 1.44x over the CTC+AR baseline (Sys. 16 \textit{vs.} 13) without a significant WER increase.
At the same decoding speed, WER reductions of up to 0.3\% absolute (6.1\% relative) are obtained over the CTC+AR baseline (Sys. 15 \textit{vs.} 13).

\noindent
\textbf{Experiments Using Beam Search: } 
Performance of AMD Decoders using beam search are in Tab. \ref{tab:beam_search}. 
The results suggest the following trends,
\textbf{(a)} the tripartite Decoder consistently speeds up decoding without significant WER increase. 
For LibriSpeech-100hr trained models, it achieves a maximum speed-up of 1.59x (Sys. 6 \textit{vs.} 1).
Models with WavLM SSL feat. and 960-hour pretraining see a 1.53x speed-up (Sys. 11 \textit{vs.} 9).
\textbf{(b)} Mixed-size decoding allows the tripartite Decoder to achieve similar or even slightly better performance than the CTC+AR baseline at the same RTF (Sys. 8 \textit{vs.} 1, Sys. 12 \textit{vs.} 9).
\textbf{(c)} For systems that exhibit similar 1-best WER (Sys. 6-8 \textit{vs.} 1, Sys. 11-12 \textit{vs.} 9), the tripartite Decoder consistently larger Oracle WERs than the CTC+AR baselines. Further analysis of this phenomenon is conducted in the next section.

\begin{figure}%
    \centering
    \subfloat[\centering Density]{{\includegraphics[width=3.8cm]{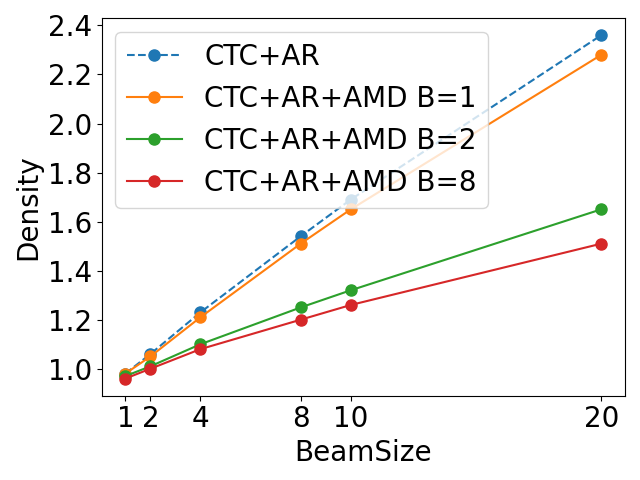} }}%
    \subfloat[\centering Oracle WER]{{\includegraphics[width=3.8cm]{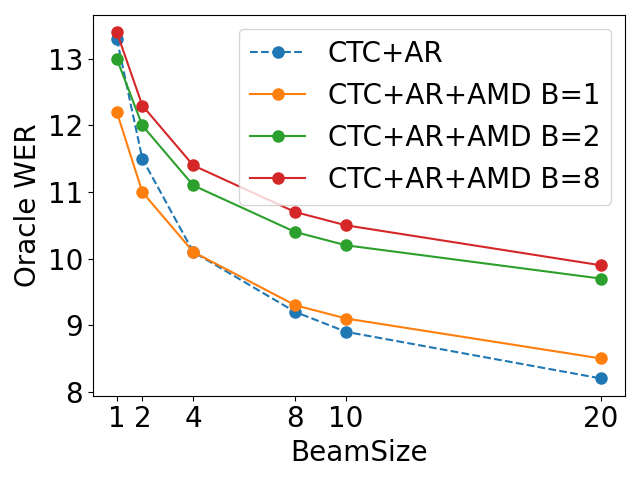} }}%
    \vspace{-2mm}
    \caption{AMD lattice density and oracle WERs computed using 100-best hypotheses over varying top-K beam size from 1 to 20.}%
    \label{fig:wero}%
    \vspace{-5mm}
\end{figure}

\noindent
\textbf{Analysis of AMD Lattice Density and Oracle WER: }
Fig. \ref{fig:wero} further explores the relationship between (a) decoding lattice density and (b) the average Oracle WER obtained from the 100-best hypotheses on "clean" and "other" test sets.
Lattice density here measures the average number of unique predictions for each token in the ground truth text, as found in the 100-best list, while Oracle WER evaluates beam search performance by identifying the single hypothesis with the lowest WER within the 100-best list.
As block size increases, the tripartite Decoder exhibits sparser lattices (lower density) and faster saturation compared to the baseline system (orange, green, and red curves \textit{vs.} dashed blue curve).
This explains why the improvements from the CTC+AR+AMD tripartite Decoder over the CTC+AR baseline when operating with greedy search under aggressive pruning of Tab. \ref{tab:greedy_search} are much larger than those found in beam search experiments of Tab. \ref{tab:beam_search}.
 This disparity on performance improvements may be due to  in the premature pruning within attention-mask blocks of AMD, as implemented in line 21, Algorithm 1. 
We plan to investigate this in future work.

\vspace{-1mm}
\section{Conclusions}
This paper proposes a non-autoregressive block-based Attention Mask Decoder (AMD) that flexibly balance performance-efficiency trade-offs. Experimental results on the LibriSpeech-100hr dataset demonstrate that the proposed tripartite Decoder incorporating the AMD module achieves a maximum decoding speed-up ratio of 1.73x over the baseline CTC-AR decoding, while while incurring no statistically significant WER increase on the ``clean'' and ``other'' test sets. When operating with the same decoding real time factors,  statistically significant WER reductions of up to 0.7\% and 0.3\% absolute (5.3\% and 6.1\% relative) were obtained over the CTC-AR baseline. Future research will focus on further refining the AMD beam search algorithm and enhancing the quality of lattice or N-best outputs.
\newpage
\section{Acknowledgements}
\label{ssec:a}
This research is supported by Hong Kong RGC GRF grant No. 14200021, 14200220, TRS grant No. T45-407/19N, Innovation \& Technology Fund grant No. ITS/218/21, National Natural Science Foundation of China 62106255, and Youth Innovation Promotion Association CAS grant 2023119.
\bibliographystyle{IEEEtran}
\bibliography{mybib}

\end{document}